\documentclass[a4paper,runningheads]{llncs}
\usepackage{graphicx}
\usepackage{amsmath}
\usepackage[utf8]{inputenc}
\usepackage{xcolor}
\usepackage{subfig}
\usepackage{hyperref}

\usepackage{booktabs}
\usepackage{xcolor}
\usepackage[shortcuts,acronym]{glossaries}

\usepackage{siunitx}
\newacronym{lasso}{LASSO}{least absolute shrinkage and selection operator}  
\newacronym{nwp}{NWP}{numerical weather prediction}
\newacronym{ml}{ML}{machine learning}
\newacronym{svr}{SVR}{support vector regression}
\newacronym{svm}{SVM}{support vector machine}
\newacronym{mlp}{MLP}{multi-layer perceptron}
\newacronym{gbrt}{GBRT}{gradient boosting regression tree}
\newacronym{mae}{MAE}{mean absolute error}
\newacronym{rmse}{RMSE}{root mean squared error}
\newacronym{mse}{MSE}{mean squared error}
\newacronym{kld}{KLD}{Kullback-Leibler Divergence}
\newacronym{pv}{PV}{photovoltaic}
\newacronym{wrt}{w.r.t.}{with respect to}
\newacronym{pdf}{PDF}{probability density function}

\begin{document}
%%% Mehrere Autoren werden durch \and voneinander getrennt.
%%% Die Fußnote enthält die Adresse sowie eine E-Mail-Adresse.
%%% Das optionale Argument (sofern angegeben) wird für die Kopfzeile verwendet.

\title{{Influences in Forecast Errors for Wind and Photovoltaic Power: A Study on Machine Learning Models}}
\titlerunning{Influences Forecast Errors}

\author{Jens Schreiber,Artjom Buschin , Bernhard Sick}

\institute{University of Kassel\\ \texttt{\{j.schreiber, artjom.buschin, bsick\}@uni-kassel.de}}

%%%\lnidoi{18.18420/provided-by-editor-02} % Falls bekannt
\maketitle
\begin{abstract}
	Despite the increasing importance of forecasts of renewable energy, current planning studies only address a general estimate of the forecast quality to be expected and selected forecast horizons.
    However, these estimates allow only a limited and highly uncertain use in the planning of electric power distribution. More reliable planning processes require considerably more information about future forecast quality. 
    In this article, we present an in-depth analysis and comparison of influencing factors regarding uncertainty in wind and photovoltaic power forecasts, based on four different \ac{ml} models.
    In our analysis, we found substantial differences in uncertainty depending on \ac{ml} models, data coverage, and seasonal patterns that have to be considered in future planning studies.
\end{abstract}
\section{Introduction}

 % An introduction to the context or background of the topic (you could include interesting facts or quotations)
 % The reason for writing about this topic
 % Definitions of any complex terminology that will be referred to throughout the assignment (note that definitions are not always necessary)
 % Introduce the main ideas that stem from your topic/title and the order in which you will discuss them?
 With the further expansion of wind and \ac{pv} energy, the power supply system will change significantly in the coming decades. The overall power supply will become more weather-dependent, and solutions must be found to ensure a robust and inexpensive power supply that maintains grid stability. Since high investments over a long period are necessary for that purpose, the expansion must be planned as precisely as possible in the long term. Simulations of different possible scenarios for the future power supply system are essential to compare different options and optimize the expansion. 

 The major challenges for the energy system transformation can mainly be traced back to two aspects: Firstly, the actual power supply of wind and solar energy plants is directly dependent on the weather and thus not directly compatible to consumption. Secondly, the expected power of the next hours and days is uncertain due to the strong dependence on the weather and must be predicted by power forecasts based on \acp{nwp}.
 
 Despite the increasing importance of forecasts for renewable power supply, current planning studies only address the forecast quality to be expected in the future for the whole of Germany based on representative forecasts (see, e.g., the dena grid study~\cite{Kohler2010}). Further, often these studies only consider a limited number of forecasts horizons. However, these estimates allow only for a limited and highly uncertain use in the planning of the electricity supply system. More reliable planning processes require considerably more information about future forecast quality.

\section{Main Contribution}
The article provides a comprehensive study on influences in forecast uncertainty which has to be taken into consideration for future planning studies. 
The articles investigates uncertainty in four types of common \ac{ml} models for wind and \ac{pv}: \Ac{lasso}, \ac{gbrt}, \ac{svr}, and \ac{mlp}. 
Models are trained to forecasts the estimated day-ahead power generation based on \ac{nwp} features as input. 
By repeated training with different test datasets, we create forecasts over the entire the datasets for later analysis. 

In the next step, we compare error distributions for each model concerning known influencing factors: Amount of training samples, forecast horizon, the terrain of wind farms, and a comparison of the uncertainties between the different \acf{ml} models. 
By comparing binned forecasts errors, e.g., for different forecast horizons, with the \ac{kld}, we measure similarities and differences of these distributions. 
This comparison allows estimating when a bin is substantially different compared to a baseline and therefore gives insights to influential factors. Further, bins are compared with the Kruskal-Wallis~\cite{Hedderich2018} hypothesis test to verify a significant difference.
The main contributions are:
\begin{itemize}
    \item We utilize common \ac{ml} models, the grid search algorithm to find optimal model parameters, and common feature engineering techniques to provide forecasts results for wind and \ac{pv} farms. By repeated training of the models on different training sets, we create forecasts for the complete dataset.
    \item For wind power forecasts, we show that the amount of training samples influences the forecasts error up to a certain threshold of data coverage (where data coverage is the proportion of the maximum number of data samples in the dataset to the actual amount of data samples within the historical data).
    \item Analyzing seasonal patterns reveals different influences for wind and \ac{pv} that are related to different weather conditions for a different season of a year. Interestingly, forecast errors of adjacent seasons are not necessarily similar to each other in \ac{pv} forecasts.
    \item The comparative study of \ac{ml} forecasts models shows, that wind forecast errors within a similar terrain are more alike than for a similar amount of training samples. This relation suggests that forecasts models are more alike when external influences are excluded.
\end{itemize}
The remainder of this article is structured as follows. In Section~\ref{sec:related_work} we detail related work. Section~\ref{sec:method} outlines evaluation measures and applied \ac{ml} models. Section~\ref{sec:experimental_evaluation} describes the experimental design and evaluation results w.r.t. data coverage, seasonal patterns, terrain, and model differences. Finally, we conclude our work and propose future work in Section~\ref{sec:conclusion}.

\section{Related Work}
\label{sec:related_work}

In current planning studies on future energy systems, considerations on the current and future uncertainty of power forecasts are only inconsiderably taken into account.  The German Dena II study~\cite{Kohler2010}, e.g., only considers forecasts error up to a horizon of two hours, neglecting that an increasing amount of renewable requires a larger forecast horizon such day-ahead forecasts. Further, the study is missing an analysis of seasonal effects and forecast model specific uncertainties.

These (mostly) missing influential factors~\cite{Yan2015} categorizes into the \ac{nwp} input data, the power curve, and the prediction algorithm. The thesis of M. Lange~\cite{Lange2003} relates forecasts error to the \ac{nwp} data. In particular, the forecast uncertainty is assessed \ac{wrt} certain meteorological situations. However, the study only employs a physical model of the power curve with error correction and spatial refinement.

In the work of~\cite{Pinson2006}, time series analysis techniques (e.g., ARIMA, ARX, Box-Jenkins) and a physical model are used to evaluate the forecasting skill. The author includes an analysis for different forecast horizons based on $\textit{R}^2$ and \ac{rmse}. Further, it contains a small subsection on the evaluation of the error distribution. 

The results of \cite{Mohrlen2004} include an analysis for time horizons between zero to nine hours and up to five days ahead. But the study is again focusing on the physical model and not considering machine learning models. Also~\cite{Holttinen2013, Ko2015} are focusing on time series analysis techniques (NARX) and (adapted) physical models, for uncertainty analysis.

More recently, in~\cite{Gensler2018} uncertainty in \ac{ml} models, such as \ac{mlp}, \ac{svr}, and an ensemble technique, are analyzed, but the thesis misses an evaluation for error distributions relating to different forecast horizons of wind power. Also, in~\cite{Reindl2017} an analysis of \ac{ml} models such as extreme gradient boosting technique, random forest, adaptive boosting, and persistence method is used to access the economic value for \ac{pv} power generation.

In~\cite{Brecl2018} uses physical and semi-physical models for developing a forecast methodology for households that do not have access to solar irradiance information and are therefore limited to discrete weather information. The results are analyzed \ac{wrt} the discrete weather features. 

An interesting approach presents~\cite{Nam2018}, in which the kriging method interpolate data with geographical properties for a location with no available data. A Na\"{i}ve bayes classifier along with a Gaussian probability distribution based on the overall data performs day-ahead forecasts of solar power based on the probability in one-hour intervals. The method is evaluated against the persistence model with \ac{mae} for different months of a year. 

The simulation in \cite{NREL2013} creates uncertainties for \ac{pv} at different time scales to evaluate the economic and reliability effect for the grid. As it is a simulation tool, it is different from \ac{ml} models. The proposed method in~\cite{Murata2018} allows for modeling the \ac{pv} uncertainty based on past observations by using multivariate normal distributions.

The literature review shows that most of the work is focusing on models relating to time series analysis techniques and physical models. Further, the reviewed articles are missing a quantified comparison between the distribution of uncertainties or are even missing an in-depth analysis of the error distribution.

% \cite{Tuononen2019} shows that the solar radiation prediction error is greater in summer than in winter, but the relative solar radiation prediction error is more or less constant throughout the year. 
% Pinson benutzt prob. vorhersagen
% Ko2015 = physical model + correction term

\section{Method}
\label{sec:method}

To evaluate influences in forecast uncertainty this section gives a summary on common \ac{ml} algorithms and present their differences. Using different \ac{ml} algorithms assures to cover a broad spectrum in forecast errors. In the final section, we summarize error measures to estimate the deviation between actual and forecasted power generation.

\subsection{Lasso}

\Ac{lasso}, also known as basis pursuit, is a linear model. 
Linear models typically provide a robust estimation, when \acp{nwp} are uncertain. 
Further linear models allow measuring the contribution of individual features through their coefficient, hence, making them highly relevant for analysis on error origin~\cite{Hastie2001}.
In contrast to other linear regression models, \ac{lasso} allows for automatic selection of essential features.
This selection is achieved by $L_1$ penalty, that effectively causes the coefficient of features to be exactly zero and hence excluding individual features.

\subsection{Support Vector Regression}
\Ac{svr} is based on the concept of \acp{svm} for classification with changes in the definition of the optimization problem. One appealing property of \acp{svm} is that the determination of parameters is locally and globally optimal due to the convex optimization~\cite{Bishop2006}. Further, by making use of the \textit{kernel-trick} original \ac{nwp} input features are transformed in a higher dimensional, even infinite dimensional, space. Transforming features into a higher-dimensional space provides features that are linearly separable~\cite{Vapnik2000}. The transformed features allow the \ac{svr} to achieve good results in many applications~\cite{Bishop2006}, making them highly relevant for the evaluation of forecast uncertainty.

\subsection{Multi-layer Perceptron}

\Acp{mlp}, and more recently deep neural networks are a common technique for regression and classification tasks. In an \ac{mlp} input features are transformed using matrix multiplication and a subsequent (mostly) non-linear transformation. The former two operations are summarized as layers and successive applications of these layers, where the output of one layer is the input to the next layers, allows us to find a good representation of the data. In the final layer, the output layer, a simple linear combination can be used for renewable energy forecast. Primarily through their capability to find good representations of the \ac{nwp} data, \acp{mlp} achieve state of the art performance in renewable power forecast~\cite{Gensler2016}. This performance makes them highly relevant for the evaluation of forecast uncertainty.

\subsection{Gradient-Boosting-Regression-Tree}
\Ac{gbrt} originate from the idea, that a combination of weak learners improves the overall performance. 
Therefore, the gradient boosting algorithm trains trees in regions of most substantial forecast error. 
The ensemble technique combines the individual trees improving the overall performance.
A single tree partitions the features space in a set of rectangles and estimates a constant forecast value for each rectangle~\cite{Hastie2001}. 
The partitioning provides an interpretable structure to explain forecast decision which is not feasible with \ac{svr} and \acp{mlp}. Further, the algorithm is not making use of any data representation techniques as with these approaches.

\subsection{Error Measures}

To assess influences in forecast uncertainty, through the forecast error, it is essential to evaluate the error with $e = y - \hat{y}$. It gives insights between the actual power generation $y$ compared to the forecasted power $\hat{y}_i$. In contrast to mean based measures, $e$ provides the most detailed view on the error; combined with a visualization of the error distribution through a histogram or a boxplot it allows to assess skewness and other statistical measures of the error distribution.
A comprehensive analysis of deterministic error measures in the field of renewable energy forecast is given in~\cite{Gensler2018}. The results can be summarized as follows
\begin{itemize}
    \item The coefficient of determination $\text{R}^2$ assesses how much of the variance in the historical power data is explained by the model. As it is only capable of evaluating the amount of linear correlation it is often used as a measure to compare different forecast techniques.
    \item To account for extreme errors of $e$, quadratic errors such as the \ac{mse} are recommend.
    \item Absolute measures such as \ac{mae} are suited for monetary evaluation criteria (linear evaluation criteria). 
\end{itemize}

In the following, we will stick to $e^2$ as it allows for comparison of overall forecast quality of the model by terms of mean (\ac{mse}) and median, especially when visualized via boxplot.

To compare distributions of errors with another we use the \ac{kld}. 
The \ac{kld} is a non-symmetric statistical measurement to determine the difference between two distributions allowing to quantify the similarity, e. g., between the error distribution from the \ac{gbrt} and the \ac{svr}.
% It is defined as
% \begin{equation}
% D_{KL}\left(P||Q\right) = \int_{-\infty}^\infty p\left(x\right) log \left(\frac{p\left(x\right)}{q\left(x\right)}\right) dx
% \end{equation}
% with $P$, $Q$ as the  distributions of forecasts and measurements and $p(x)$,  $q(x)$ as their \acp{pdf}. Due to the non-symmetrical behavior, both $D(P||Q)$ and $D(Q||P)$ are calculated and added together.
% One interpretation of the  KLD is as the information gain achieved by replacing distribution $Q$ with $P$.

\section{Experimental Evaluation}
\label{sec:experimental_evaluation}
In the following, we provide analysis on error distributions from different \ac{ml} models and measure their similarity to another for wind and \ac{pv}. 
By estimating the \ac{kld} between distributions for different (external) factors we get insights on how they relate to another.
Therefore, we first give details on the model training and the two datasets.
The first study estimates influences caused by a limited amount of training samples for wind power forecasts.
Results are evaluated \ac{wrt} the data coverage, where data coverage refers to the proportion between the maximum number of data samples to the actual amount of data samples within the dataset.
In the next section, we analyze seasonal influences such as the hour of the day or season of the year for wind and \ac{pv} as well as terrain specific influences in the wind dataset.
As results for \ac{pv} models suggest that there are strong seasonal patterns to consider - that are less present for wind models - we limit the final analysis to the WindFarm dataset.  Limiting these and other external influences allow to compare the error distributions of the different power forecasting models.

% After analyzing external influences to the error distributions, in this section, we are interested in comparing the similarity between the \ac{ml} models. 
% As results for \ac{pv} models suggest that there are strong seasonal patterns to consider - that are less present for wind models - we limit further analysis to the WindFarm dataset.

\subsection{Design of Experiment}
For the following two datasets we train the \ac{lasso}, \ac{svr}, \ac{mlp}, and \ac{gbrt} to forecast the power generation.

% https://git.ies.uni-kassel.de/jens/prophesy/blob/master/examples/forecast_pv.ipynb
\textbf{Solar Farm Dataset:}
The \textit{SolarFarm} dataset consists of $114$ PV facilities in Germany. 
Their installed nominal power ranges between 7.2\SI{}{\kilo\watt} and 12573\SI{}{\kilo\watt}. 
% The PV facilities range from PV panels installed on rooftops to fully fledged solar farms. 
% All facilities are in Germany. 
The dataset has a three-hour resolution and is recorded from the beginning of 2016 to the end of April 2017 resulting in a maximum of $3880$ data points.
In total the dataset has 51 input features as input. 
Features with correlation to the power generation (e.g., sun position, solar height, clear sky, and radiation)
are shifted in time by three hours to take future and past effects of the weather into account for prediction.
% \todobox{Sun position theta z, and sun position extra terr, sun position solar height, clear sky direct, clear sky diffuse, clear sky global, relative humidity, net solar radiation, solar radiation direct, solar radiation diffuse }, with correlation to the power generation, 
% https://git.ies.uni-kassel.de/jens/prophesy/blob/master/examples/forecast_wind.ipynb
% \noindent

\textbf{Wind Farm Dataset}:
The \textit{WindFarm} dataset contains the power generation taken from $54$ wind farms that are distributed throughout Germany. 
These values were recorded hourly over two years (2016 and 2017) resulting in a maximum of $17520$ data points. 
The dataset contains information about the terrain of each farm (flatland, forest, and offshore). In total the dataset has $7$ \ac{nwp} features as input. 
Features of wind speed and wind direction influencing the power generation~\cite{Jens2018} are time-shifted by two hours to take future and past effects of the weather into account for prediction.

Both datasets were manually filtered to remove outliers, e.g., caused by maintenance. 
Depending on the number of outliers and maintenance the amount of data coverage ranges between $50$ and $100$ percent for wind data \ac{wrt} recorded period, where data coverage refers to the proportion between the maximum number of data samples to the actual amount of data samples within the dataset.
The data coverage for \ac{pv} is mostly above $90$\%.

To compare forecast errors, we normalize the generated power by the maximum power generation. 
Input features are standardized for zero mean and unit variance based on the training dataset in each run.
We optimize each model through a grid search on the validation dataset. 
To make the best use of the full data range, we use different runs of the experiment to shift the test data throughout the recorded period: Six months for the wind and four months for the \ac{pv} dataset resulting in four runs for each dataset.
In each run, the remaining data is used for training ($80$\%) and validation ($20$\%).  
After completing all training runs, combing predictions from all test datasets provides an \textit{evaluation dataset} for estimating influences in the complete period of the original data. 
To account for extreme errors and measure the quality between a single forecast and the historical power we use the squared error. 
We fit distributions of the squared error with the $\tilde{\chi}^2$ distribution to compare them with the \ac{kld}.
% %  for the following parameters:
% \begin{itemize}
%     \item \textbf{\Ac{svr}}: C($1e-2$-1e3), gamma($1e-2$-$1e3$), tolerance=1e-4,kernel=rbf.
%     \item \textbf{\Ac{lasso}}: Alpha ($0.1$ - $1$), max iter ($1000$-$4000$).
%     \item \textbf{\Ac{gbrt}}: max depth ($12$ -$20$,) min samples split ($2$-$128$),n estimators ($1200$-$2000$),learning rate  ($.05$-$0.2$).
%     \item \textbf{\Ac{gbrt}}: max iter = $10000$, early stopping=True/False , learning rate ($1e-2$-$1e-6$),  learning rate = constant / adaptive, solver=adam, tolerance=$1e-4$
%     % hidden_layer_sizes':[ (100, 50, 30, 15, 5), (50, 100, 200, 50), (20, 10, 5) ],
% \end{itemize} 

\subsection{Influence of the Amount of Training Data}

The digitalization of the current and future energy market will provide an increasing amount of training data. 
To determine the extent to which the amount of training data influences the forecast error we analyze it in this section.

\begin{figure}[!tb]
    \centering
    \includegraphics[width=0.63\textwidth]{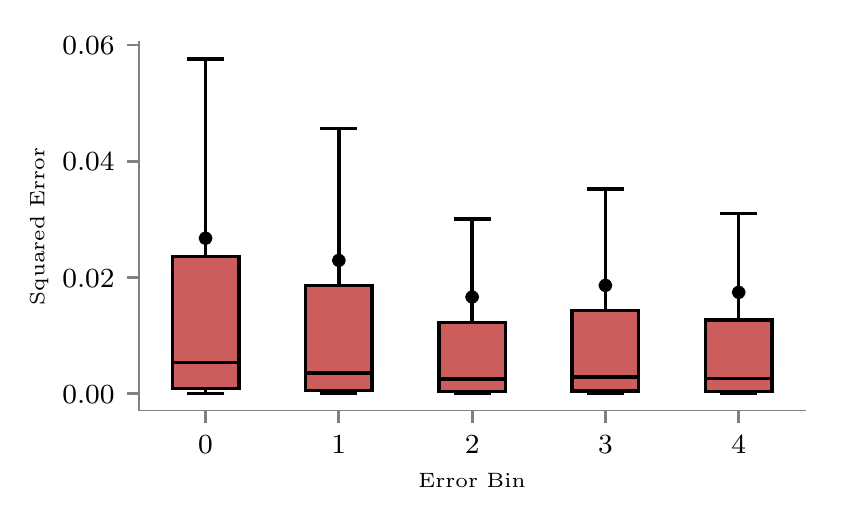}
    \caption{Boxplot on squared error for bins of data coverage for the WindFarm dataset based on the \ac{mlp} model. From left to right the data coverage is as follows: $50-60\%$, $60-70\%$, $70-80\%$, $80-90\%$, and $90-100\%$. The mean of the error is visualized as a dot.}
    \label{fig_data_cov_ann}
\end{figure}

Therefore we estimate the data coverage of a farm in percent compared to the maximum number of data points. 
It turns out that the data coverage in \ac{pv} farms is consistently above $92\%$ except for one farm, respectively we do not consider it in further analysis.
However, the data coverage from wind farm range between $49$ and $99\%$ allowing for clustering them in ten percent steps, see Figure~\ref{fig_data_cov_ann}.
As the size of the test dataset is constant in each run, the size of the training data is as well, respectively, the data coverage is directly linked to the amount of training data and will be treated equally in the following.

Figure~\ref{fig_data_cov_ann} shows the relation between the number of training samples and the error:
With the increasing amount of data, the median as well as the mean decrease. 
The spread of the error is similar for bin two, three, and four. 
Bin zero and one have a broader spread of the error.
The decreasing mean, median, and spread show that there is a relation between the amount of data for training and the forecast error.

% To verify a significant difference between these bins, we compare the error distribution of each bin with bin four measuring the dissimilarity to the maximum number of data points. 
% For each model and dataset (except for bin $3$ for GBRT in the WindFarm dataset)  the Kruskal-Wallis hypothesis test, rejects the null hypothesis that the median is of the bins are equal motivating their difference.
% Each bin of each dataset and model is compared to bin four. 
To verify a significant difference between these bins, we compare them with the \ac{kld} and the Kruskal-Wallis hypothesis test. Kruskal-Wallis hypothesis shows that the forecast error for all \ac{ml} models and all bins of data coverage are significantly different at a significance level of $\alpha=5\%$.
The exemplary results in Table~\ref{tbl_kld_error_bins}, highlight the previous observation: A decreasing data coverage, causes an increased spread, median, and mean resulting in larger values of the \ac{kld}, e.g., when comparing bin zero with bin four.
Bin two, three, and four are quite similar to each other nonetheless.
\begin{table}[!tb]   
\centering
\footnotesize
\begin{tabular}{lSSSSS}
    \toprule
    {Bins} &    {0} &         {1} &         2 &         3 &         4 \\
    \midrule
    0 &  0.0 &  0.002601 &  0.043580 &  0.036882 &  0.065464 \\
    1 &   &  0.000000 &  0.067652 &  0.059214 &  0.094466 \\
    2 &   & &  0.000000 &  0.000277 &  0.002189 \\
    3 &   &  &   &  0.000000 &  0.004023 \\
    % 4 &  0.0 &  0.000000 &  0.000000 &  0.000000 &  0.000000 \\
    \bottomrule
\end{tabular}
\caption{\Ac{kld} measuring the similarity between different amounts of data coverage for the error of the GBRT model on the WindFarm dataset. From one to five the data coverage is as follows: $50-60\%$, $60-70\%$, $70-80\%$, $80-90\%$, and $90-100\%$.}
\label{tbl_kld_error_bins}
\end{table}    

As expected there is a relation between the amount of data available and the forecast error. 
With an increasing amount of available data, the \ac{ml} model tends towards a minimum error, the \ac{nwp} input data probably cause that.

\subsection{Influences by Seasonal Patterns and Terrain}

Seasonal influences that are present in seasons of a year or hours of a day are well known. 
Nonetheless, there is limited research on how these patterns affect the error distribution in wind and \ac{pv} forecast based on \ac{ml} techniques.
More common is the analysis of forecast error \ac{wrt} their terrain, which this section also covers. 

% This section covers influences that are related to a specific terrain of a wind farm or are caused by seasonal patterns for wind as well as \ac{pv}. 
In the following, we address season of a year and the hour of the day. 
In terms of \ac{pv}, the hour of the day has two meanings. 
First, due to the daily pattern of the sun, we can observe patterns within the power generation. 
Second, with the rising time of the day, the forecast horizon of the \ac{nwp} model increases (as the so-called \ac{nwp} model run typically originates from 12 UTC). 
As the horizon increases, the error of the weather forecast increases and respectively that of the power forecast model.
The latter also holds for wind power forecasts.
\begin{figure}[!tb]
    \centering
    \includegraphics[width=0.63\textwidth]{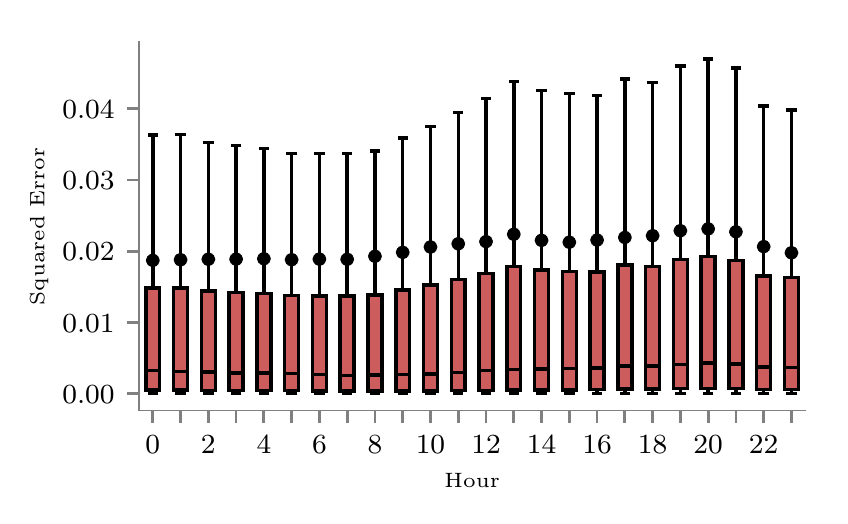}
    \caption{Boxplot on squared error for bins of the hour of the day for the WindFarm dataset based on the \ac{mlp} model. The mean of the error is visualized as a dot.}
    \label{fig_hour_ann_wind}
\end{figure}

In the sample boxplot, Figure~\ref{fig_hour_ann_wind} for wind errors we can observe this pattern. 
The median and mean errors for different hours of the day do not increase drastically due to the absence of seasonal weather patterns in the wind; detailed observations exist when measuring similarity through \ac{kld} in Table~\ref{tbl_kld_hour}.
The errors at the end of the day are more similar to another by means of the \ac{kld}, compared to the origin of the weather forecast due to the increased forecast error of the \acp{nwp}. 
Nonetheless, all errors, when comparing different hours of the day, are significantly different in the Kruskal-Wallis hypothesis test ($\alpha=5\%$) except: Four cases in the linear model, one for the \ac{mlp}, and two comparisons for the \ac{gbrt} model.

\begin{table}[!bt]  
    \centering
    \footnotesize
    \begin{tabular}{lSSSSSSSS}
    \toprule
    {Hour} &    0 &         3 &         6 &          9 &          12 &          15 &          18 &          21 \\
    \midrule
    0 &  0.0 &  0.000428 &  0.056646 &  1.905154 &  7.618584 &  5.149946 &  0.830981 &  0.437450 \\
    3 &  &   0.000000 & 0.067008 &  1.975407 &  7.815359 &  5.293545 &  0.873054 &  0.466807 \\
    6 &  &   &   0.000000&  1.211826 &  5.635978 &  3.707136 &  0.430930 &  0.172545 \\
    9 &  &   &   &  0.000000 &  1.002574 &  0.448542 &  0.167518 &  0.422723 \\
    12 &  &   &   &   &  0.000000 &  0.094223 &  2.189676 &  3.168283 \\
    15 &  &   &   &   &   &  0.000000 &  1.243763 &  1.931795 \\
    18 &  &   &   &   &   &   &  0.000000 &  0.054416 \\
    % 21 &  0.0 &   &   &    &    &    &    &    \\
    \bottomrule
    \end{tabular}
    \caption{\Ac{kld} measuring the similarity between hours of a day for the error of the \ac{mlp} model on the PVFarm dataset.}
    \label{tbl_kld_hour}
\end{table}    

\begin{figure}[!bt]
    \centering
    \includegraphics[width=0.63\textwidth]{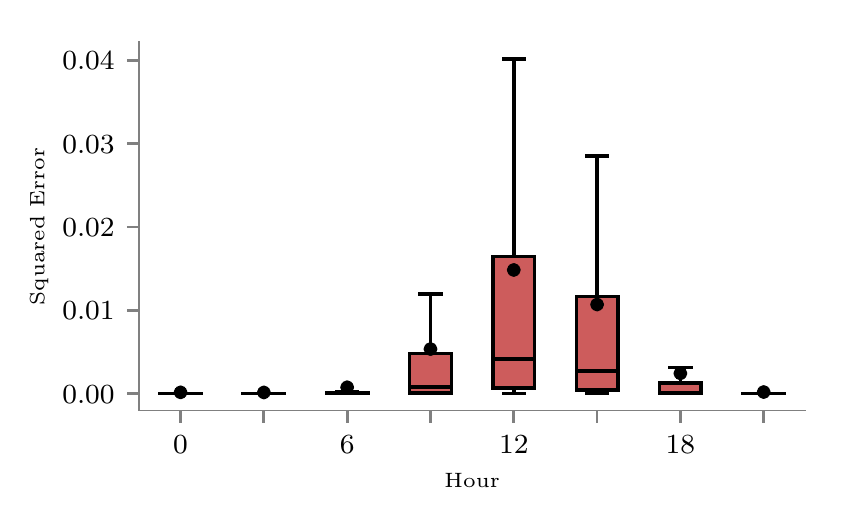}
    \caption{Boxplot on squared error for bins of an hour of the day for the SolarFarm dataset based on the \ac{gbrt} model. The mean of the error is visualized as a dot.}
    \label{fig_hour_gbrt_pv}
\end{figure}

For \ac{pv} however, see Figure~\ref{fig_hour_gbrt_pv}, we can observe a strong seasonal pattern in the error distributions for different hours of the day. 
This observation holds even more true when estimating the \ac{kld} resulting in substantial large values when comparing $12$ with $0$ o'clock, see Table~\ref{tbl_kld_hour}. 
This seasonal pattern is to expected, as during the night there is no power generation, and respectively the difference in the error distribution is notable when compared to the day.
Compared to wind, there are also more considerable differences in the error distributions during the daytime. 
The daily pattern of the sun causes these differences that result in different error distributions.
Again, all errors, when comparing different hours of the day, are significantly different in the Kruskal-Wallis hypothesis test ($\alpha=5\%$) except: Four cases in the \ac{svr} model and one case for the \ac{mlp} model.

In the analysis for different seasons of a year for wind, we observe that in the third season all models and datasets have the lowest median, mean, and spread of the error for wind. 
In other seasons of the year, extreme weather conditions are more common, causing larger error values.
The \ac{kld}, see Table~\ref{tbl_kld_season_wind}, confirms our intuition, that error distributions for seasons close to another are more similar to another than those far away.

\begin{table}[!tb]   
    % \caption{\Ac{kld} measuring the similarity between seasons of a year for wind and \ac{pv}.}
    \centering
    \footnotesize
    \sisetup{round-precision=2}
    \begin{minipage}{.45\textwidth}
            
            \begin{tabular}{lSSSSS}
            \toprule
            {Season} &     1 &         2 &         3 &         4 \\
            \midrule
            1 &  0.0 &  0.015981 &  0.185749 &  0.100555 \\
            2 &   &  0.000000 &  0.091524 &  0.036002 \\
            3 &   &  &  0.000000 &  0.012541 \\
            % 3 &  0.0 &  0.000000 &  0.000000 &  0.000000 \\
            \bottomrule
        \end{tabular}
        \caption{\Ac{kld} measuring the similarity between seasons of a year for the error of the SVR model on the WindFarm dataset. Season one equals winter, season two equals spring, season three equals summer, and season four equals autumn.}
        \label{tbl_kld_season_wind}  
    \end{minipage}  
    \hspace{2em}
    \centering
    \begin{minipage}{.45\textwidth}
        \begin{tabular}{lSSSSS}
        \toprule
        {Season} &    1 &         2 &         3 &         4 \\
        \midrule
        1 &  0.0 &  0.184669 &  3.214814 &  0.084939 \\
        2 &   &  0.000000 &  1.579525 &  0.018544 \\
        3 &   &  &  0.000000 &  2.017891 \\
        % 3 &  0.0 &  0.000000 &  0.000000 &  0.000000 \\
        \bottomrule
    \end{tabular}
    \caption{\Ac{kld} measuring the similarity between seasons of a year for the error of the \ac{svr} model on the SolarFarm dataset. Season one equals winter, season two equals spring, season three equals summer, and season four equals autumn.}
    \label{tbl_kld_season_pv}   
    \end{minipage} 
\end{table}

Contrary to wind forecasts, \ac{pv} models have more substantial errors in the third season. 
In other seasons of the year, the different position of the sun causes a different amount of direct and diffuse radiation making it the forecast model easier to forecast the power generation. 
For instance, the solar radiation (direct and diffuse) is the smallest in season one in the dataset.
The analysis of the \ac{kld} in Table~\ref{tbl_kld_season_pv} suggest that the difference of uncertainty is significant even for seasons of the year that are close to another. 
These differences are caused by the larger magnitude of the forecast error, especially in the third season.
Only when comparing season one and three for errors of the \ac{gbrt} model on the WindFarm dataset the Kruskal-Wallis ($\alpha=0.05$) estimates no significant difference.

\begin{figure}[!tb]
    \centering
    \includegraphics[width=0.63\textwidth]{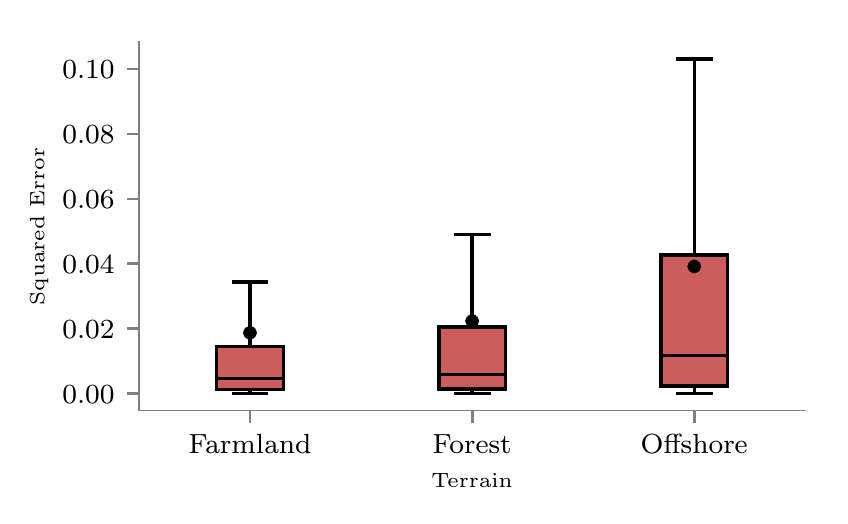}
    \caption{Boxplot on squared error for different terrains for the WindFarm dataset based on the \ac{gbrt} model. The mean of the error is visualized as a dot.}
    \label{fig_terrain_gbrt}
\end{figure}

The analysis of the terrain in Figure~\ref{fig_terrain_gbrt} shows that the smallest errors are present for parks located in a farmland terrain. 
All errors, when comparing the different terrains, are significantly different in the Kruskal-Wallis hypothesis test ($\alpha=5\%$).
Note that the terrains have a varying amount of farms. 
Farmland has  $37$, the forest has $11$, and offshore includes four farms. 
Interestingly, when measuring the similarity, the error distribution of offshore farms is closer to the farmland, than farmland to the forest employing the \ac{kld}.
This smaller \ac{kld} might be due to more complex weather conditions in the forest and offshore terrain. 
For instance, turbulence on the sea might be similarly present in forests (that are often also elevated) causing a similar uncertainty distribution.

%  GBRT Terrain differences
% \begin{tabular}{lrrr}
%     \toprule
%     {} &    Farmland &         Forest &         Offshore \\
%     \midrule
%     Farmland &  0.0 &  0.019207 &  0.142861 \\
%     Forest  &  0.0 &  0.000000 &  0.056564 \\
%     % Offshore &  0.0 &  0.000000 &  0.000000 \\
%     \bottomrule
% \end{tabular}

% \begin{table}   
% \centering
% \begin{tabular}{lSSSSSS}
%     \toprule
%     {Hour} &    0 &         4 &             8 &         12 &         16 &         20 \\
%     \midrule
%     0 &  0.0 &  0.167821 &  1.670420e-01 &  0.031301 &  0.062153 &  0.041654 \\
%     4 &  0.0 &  0.000000 &  8.690260e-07 &  0.053206 &  0.025074 &  0.041394 \\
%     8 &  0.0 &  0.000000 &  0.000000e+00 &  0.052774 &  0.024779 &  0.041014 \\
%     12 &  0.0 &  0.000000 &  0.000000e+00 &  0.000000 &  0.005182 &  0.000732 \\
%     16 &  0.0 &  0.000000 &  0.000000e+00 &  0.000000 &  0.000000 &  0.002018 \\
%     20 &  0.0 &  0.000000 &  0.000000e+00 &  0.000000 &  0.000000 &  0.000000 \\
%     \bottomrule
%     \end{tabular}
%     \caption{\Ac{kld} measuring the similarity between different hours day for the error of the SVR model on the WindFarm dataset.}
% \label{tbl_kld_hour}
% \end{table}       

Conclusively, we showed similarity and dissimilarity in seasonal and terrain specific patterns. 
Interestingly, the difference in error distribution is one of the largest for different seasons of the year for wind and \ac{pv}. 
Wind errors are more significant in the winter and autumn, while \ac{pv} models have larger values in the spring and summer time. 
Finally, the uncertainty distribution in offshore terrain is like that of the forest.

\subsection{Influences by Models}

After analyzing external influences to the error distributions, in this section, we are interested in comparing the similarity between the \ac{ml} models. 
As results for \ac{pv} models suggest that there are strong seasonal patterns to consider - that are less present for wind models - we limit further analysis to the WindFarm dataset.

In previous results from wind models, we show that the error is the smallest for the farmland terrain and when training the \ac{ml} model on a data coverage between $90$ to $100\%$.

Limiting the analysis to the smallest errors gives us insights, more similar to an absence of external influences.
As those influences must have a smaller effect on the error distribution compared to distribution with larger mean, median, and spread. 
Ultimately, allowing to access the differences in the \ac{ml} models and not those caused by external influences.

In both analysis we observe that \ac{gbrt} achieves the smallest error, \ac{svr} the second, \ac{mlp} the third smallest and \ac{lasso} has the most substantial error.
Table~\ref{tbl_kld_max_data_cov_model} and \ref{tbl_kld_terrain_model} summarize the the difference in their error distributions for the experiment with maximum data coverage and the farmland terrain.

\begin{table}[!tb]
\centering
\footnotesize
% data coverage
\begin{tabular}{lSSSS}
    \toprule
    {} &    GBRT &         LASSO &         SVR &         MLP \\
    \midrule
    GBRT &  0.0 &  1.527968 &  0.047417 &  0.142267 \\
    LASSO &  &  0.000000 &  0.975012 &  0.663395 \\
    SVR &   &   &  0.000000 &  0.024907 \\
    % MLP &  0.0 &  0.000000 &  0.000000 &  0.000000 \\
    \bottomrule
\end{tabular}
\caption{\Ac{kld} measuring the similarity between \ac{ml} models with data coverage between $90$ and $100\%$.}
\label{tbl_kld_max_data_cov_model}
\end{table}  
  
\begin{table}[!tb]
\centering
\footnotesize
% terrain
\begin{tabular}{lSSSS}
        \toprule
        {} &    GBRT &         LASSO &        SVR &         MLP \\
        \midrule
        GBRT &  0.0 &  1.460152 &  0.04509 &  0.148716 \\
        LASSO &   &  0.000000 &  0.93522 &  0.608504 \\
        SVR &   &   &  0.00000 &  0.029429 \\
        % MLP &  0.0 &  0.000000 &  0.00000 &  0.000000 \\
        \bottomrule
\end{tabular}
\caption{\Ac{kld} measuring the similarity between \ac{ml} models for a farmland terrain.}
\label{tbl_kld_terrain_model}
\end{table}          

Results suggest that distributions within a terrain are more similar to another than within maximum data coverage caused by the relation that specific weather conditions, are individual for different terrains, resulting in terrain specific forecast errors. 
Nonetheless, estimates of the Kruskal-Wallis hypothesis test ($\alpha=5\%$) shows that they are still substantially different.
\section{Conclusion and Future Work}
\label{sec:conclusion}
In this article, we presented an in-depth analysis and comparison for influencing factors of uncertainty in wind and \ac{pv} power forecasts based on four different \ac{ml} models. 
In our analysis, we found substantial influences and differences between compared bins of uncertainty revealing the need to consider them in future planning studies.

For instance, the study reveals strong seasonal patterns in the uncertainty for wind and \ac{pv}.
For wind power forecasts, neighboring seasons and hours are similar to each other. 
For seasonal patterns within a year, these forecasts will benefit from optimizing \ac{nwp} forecasts for extreme weather situations that cause substantial errors in the winter time. 
Due to significantly larger errors for the third season adjacent seasonal bins in \ac{pv} forecasts are not necessarily similar to each other. 
Similar results are obtained for daily patterns.
For daily patterns, we recommend to use \ac{nwp} forecasts that are closer to the time (noon) of most substantial error. 

By analyzing the relation between the amount of training data and the uncertainty we showed that models improve when using additional data up to a data coverage of about $70$\%.
Reducing this error further is, e.g., possible with deep learning models that have a higher capacity to learn the relation between \ac{nwp} features. 
However, even with an increasing amount of data, the minimum forecast error will be limited to that error caused by the \ac{nwp}.

The study reveals that after minimizing external influences, differences in the uncertainty distributions from the four \ac{ml} models are still present motivating the need to consider the underlying forecast model in future planning studies.

In the future, we aim to investigate how \textit{transfer learning} can be utilized to reduce forecast uncertainty when limited data is available.

% \todobox{Own opinion: Provide an educated discussion showing your opinion}
% \todobox{Benefits: Summarize whoe benefits from your work}
% \todobox{On the basis of your article, which follo-up activities do you suggest and why?}

% \begin{itemize}
%     \item mehr data besser , wird sich in zukunft durch digitalisierung verbessern 
%     \item bereits gefilter --> datenqualitaet wichtig --> bessere filter in zulunft(denoising ae)
% \end{itemize}

\small
\textbf{Acknowledgement}
This work was supported within the project Prophesy (0324104A) funded by BMWi (Deusches Bundesministerium für Wirtschaft und Energie / German Federal Ministry for Economic Affairs and Energy).

% \printbibliography
\vspace{-0.5em}
\bibliographystyle{unsrt}
\bibliography{references.bib}

\end{document}